\documentclass{aastex}
\usepackage{spr-astr-addons}
\usepackage{url}
\usepackage{multirow}

\usepackage{hyperref}
\usepackage{graphicx}

\def\aap{Astron.~\& Astrophys.}

\RequirePackage{color}

\begin{document}

\title{The analysis of effective galaxies number count for Chinese Space Station Optical Survey(CSS-OS) by image simulation}

\author{Xin Zhang\altaffilmark{1,2}} 
\affil{zhangx@nao.cas.cn}
\and 
\author{Li Cao\altaffilmark{2}}
\affil{caoli@nao.cas.cn}
\and
\author{Xianmin Meng\altaffilmark{2}}
\affil{mengxm@nao.cas.cn}

\altaffiltext{1}{University of Chinese Academy of Sciences, Beijing 100049, China}
\altaffiltext{2}{Key Laboratory of Space Astronomy and Technology, National Astronomical Observatories,Chinese Academy of Sciences, Beijing 100012, China}

\begin{abstract}
The Chinese Space Station Optical Survey (CSS-OS) is a mission to explore the vast universe. This mission will equip a 2-meter space telescope to perform a multi-band NUV-optical large area survey (over 40$\%$ of the sky) and deep survey ($\sim 1\%$ of the sky) for the cosmological and astronomical goals. Galaxy detection is one of the most important methods to achieve scientific  goals. In this paper, we evaluate the galaxy number density for CSS-OS in i band (depth, i $\sim$ 26 for large area survey and $\sim$ 27 for the deep survey, point source, 5$\sigma$) by the method of image simulation. We also compare galaxies detected by CSS-OS with that of LSST (i $\sim$ 27, point source, 5$\sigma$). In our simulation, the HUDF galaxy catalogs are used to create mock images due to long enough integration time which meets the completeness requirements of the galaxy analysis for CSS-OS and LSST.  The galaxy surface profile and spectrum are produced by the morphological information, photometric redshift and SEDs from the catalogs. The instrumental features and the environmental condition are also considered to produce the mock galaxy images. The galaxies of CSS-OS and LSST are both extracted by SExtractor from the mock i band image and matched with the original catalog. Through the analysis of the extracted galaxies, we find that the effective galaxy number count is $\sim$ 13 arcmin$^{-2}$, $\sim$ 40  arcmin$^{-2}$ and $\sim$ 42  arcmin$^{-2}$ for CSS-OS large area survey, CSS-OS deep survey and LSST, respectively. Moreover,  CSS-OS shows the advantage in small galaxy detection with high spatial resolution, especially for the  deep survey: about 20$\%$ of the galaxies detected by CSS-OS deep survey are not detected by LSST, and they have a small effective radius of r$_e < $ 0.3".

\end{abstract}


\maketitle



\section{Introduction}

The implementation of the large astronomical survey project has led to the revolutionary development of astronomy. Many new astronomical phenomena have been found relying on these large survey projects. The Sloan Digital Sky Survey (SDSS) (\cite{2000AJ....120.1579Y}) created a precedent for digital sky surveys which not only provided large area and more accurate astronomical measurement data but also developed a lot of modern astronomical data processing methods. Consequently, a number of sky survey projects are planned for the success of the SDSS mode. Euclid (\cite{2011arXiv1110.3193L}) is a 1.2-meter space-based telescope with optical and IR wide field imagers and slitless spectroscopy, proposed by ESA. The Wide Field Infrared Survey Telescope (WFIRST) (\cite{2012arXiv1208.4012G}, \cite{2013arXiv1305.5422S}) is the most sensitive infrared space-based telescope with a 2.4-meter aperture for its image and slitless spectrum observation whose bandwidth ranges from 0.76um to 2um. The Large Synoptic Survey Telescope (LSST) (\cite{2009arXiv0912.0201L}) uses a ground-based telescope with 8.4-meter primary mirror for mapping the color sky with 6 filters from 300nm to 1000nm. LSST will cover 18,000 deg$^2$ and its final (coadded) survey depth, i $\sim$ 27 AB magnitude. All of these planned or ongoing sky survey projects require large continuous sky coverage and deeper observation. From the statistical result of large observation samples, these sky survey projects will measure the expansion history of the universe and the growth rate of the structure using Baryonic Acoustic Oscillations, redshift-space distortions and clusters of galaxies, and will also look for the dark matter and the dark energy from the phenomenon of gravity lensing. 
·
Chinese Space Station Optical Survey (CSS-OS) (\cite{2011SSPMA..41.1441Z}, \cite{2018cosp...42E3821Z}) is the first large space optical astronomical project of China established by the Space Application System of the China Manned Space Program. CSS-OS will equip a 2-meter space-based telescope with high spatial resolution, R$_{80}$ (the radius of 80\% energy for an image of a point source) $\sim$ 0.15", and large Field of View (FOV) $\sim$ 1.1 deg$^2$. This project will execute both photometric imaging and slitless spectroscopic observations simultaneously. There are 7 bands and 3 gratings on the focal plane whose wavelength ranges from 250nm to 1000nm. \cite{2018MNRAS.480.2178C} introduce  the details on the filters of this project which include the peak wavelength and the widths of all filters. CSS-OS plans to cover over 40$\%$ of the sky with large area survey (coadded depth, i $\sim$ 26 AB magnitude) that distributes in the mid-high galactic latitude sky areas to probe for dark matter and dark energy, baryonic acoustic oscillation, galaxy evolution and so on. In addition, there are $\sim 1\%$ of the sky as deep survey areas (coadded depth, i $\sim$ 27 AB magnitude) by CSS-OS (hereafter, all descriptions and analysis are for large area survey if there is no description with “deep”). The main features of CSS-OS are shown in Table~\ref{tab:CSSOS_info}.

Table~\ref{tab:prjs_comp} shows the main property of some planning survey projects. From this table, we can find that CSS-OS has advantages in many aspects, such as cover areas, depth, wavelength range, and filter number. CSS-OS, as a space-based project, is similar to the ground-based project - LSST in the  aspect of the survey design. Both of CSS-OS and LSST have the plan to cover large sky areas and designs a wide spectrum coverage. Due to lack of the impact of atmosphere, there is an additional near-ultraviolet filter (NUV) for CSS-OS comparing to LSST. The Galaxy Evolution Explorer (GALEX) (\cite{2014Ap&SS.354..103B}) has imaged sky in the Ultraviolet and provide the roadmap for the future UV survey project. Comparing with GALEX, CSS-OS is lacking the FUV domain, but the depth of NUV (shown in Table~\ref{tab:CSSOS_info}) is deeper than that of GALEX. Therefore, we anticipate that CSS-OS will have new discoveries about QSOs, star-forming galaxies, nebulae, and the interstellar medium. Except for the NUV filter, the wavelength range and band number of CSS-OS is almost the same as that of LSST. However, since LSST is a ground-based project and CSS-OS is a space-based project, the characteristics of observed targets from them must be different for the data from the same wavelength range. In general, the space-based telescopes have high spatial resolution and low noise, so they can get a sharper view. Meanwhile, for the ground-based telescopes, image quality is still a problem due to the atmospheric disturbance, though the adaptive optics technology has been widely used. The seeing of LSST (FWHM of the image of a point source) is about 0.7" while the PSF of CSS-OS is about 0.15" (the radius of 80\% energy for an image of a point source), which will result in the discrepancy of observed galaxies between CSS-OS and LSST. The galaxy samples are the fundamentals of astronomical researches. Therefore, we simulate galaxy images, obtain effective galaxy number density for both CSS-OS and LSST from mock images, and analyze the discrepancy of the galaxies observed by the two projects in this paper.

\begin{table*}
\caption{The main features of CSS-OS}
\label{tab:CSSOS_info}
\begin{tabular}{l|l|l|l|l|l|l|l|l}

\tableline
Aperture & \multicolumn{8}{|l}{2 m}\\
\tableline
Field-of-View & \multicolumn{8}{|l}{1.1 deg$^2$}\\
\tableline
Image quality & \multicolumn{8}{|l}{radius encircling 80$\%$ energy 0.15''}\\
\tableline
Image pixel angular size & \multicolumn{8}{|l}{0.075''}\\
\tableline
Wavelength range & \multicolumn{8}{|l}{255 - 1000 nm}\\
\tableline
Orbit & \multicolumn{8}{|l}{Low Earth orbit}\\
\tableline
\multicolumn{9}{c}{Survey Capticity}\\
\tableline
Large area survey & \multicolumn{8}{l}{17500 deg$^2$}\\
\tableline
Deep survey & \multicolumn{8}{l}{400 deg$^2$}\\
\tableline
\multicolumn{9}{c}{Image Sensitivity (point source with 5$\sigma$ detection)}\\
\tableline
&NUV & u & g & r & i & z & y & Description\\
\tableline
Large area survey &25.4 & 25.4 & 26.3 & 26.0 & 25.9 & 25.2 & 24.4 & exposure time = 150 s $\times$ 2\\  
\tableline
Deep survey & 26.7 & 26.7 & 27.5 & 27.2 & 27.0 & 26.4 & 25.7 & exposure time = 250 s $\times$ 8\\
\tableline
\end{tabular}
\end{table*}

\begin{table*}
\caption{The comparison between some astronomical projects}
\label{tab:prjs_comp}
\begin{tabular}{l|l|l|l|l|l|l|l|l}
\tableline
\multicolumn{2}{l|}{} & Aperture & FoV & pixel size & area & wavelength range & filter  & depth \\

\multicolumn{2}{l|}{} & (m) & (deg$^2$ )  & (") & (deg$^2$) & (nm) & number &  \\
\tableline
\multicolumn{2}{l|}{CSS-OS} & 2 & 1.1 & 0.075 & 17500 & 255 - 1000 & 7 & see in Table \ref{tab:CSSOS_info} \\
\tableline
\multirow{2}{*}{Euclid\tablenotemark{a}}  & VIS & \multirow{2}{*}{1.2} & 0.56  & 0.1 & 15000 & 550 - 920 & 1 &  24.5 (10$\sigma$ extended source)\\
\cline{2-2}
\cline{4-9}
 &NISP & & 0.55 & 0.3 & 15000 & 920-2000 & 3 & 24 (5$\sigma$, point source) \\
\tableline
\multicolumn{2}{l|}{\multirow{2}{*}{Wfirst\tablenotemark{b}}} & \multirow{2}{*}{2.4}  & \multirow{2}{*}{0.28} & \multirow{2}{*}{0.11} & \multirow{2}{*}{2000} & \multirow{2}{*}{927-2000} & \multirow{2}{*}{4} & Y : 26.7, J : 26.9 H : 26.7,  \\
\multicolumn{2}{l|}{}& & & & & & & F184 : 26.2 (5$\sigma$, point source)\\
\tableline
\multicolumn{2}{l|}{\multirow{2}{*}{LSST\tablenotemark{c}}} & \multirow{2}{*}{6.7} & \multirow{2}{*}{9.6} & \multirow{2}{*}{0.2} & \multirow{2}{*}{18000} & \multirow{2}{*}{320-1080} & \multirow{2}{*}{6} & u: 26.3, g: 27.5, r: 27.7 i: 27.0,  \\
\multicolumn{2}{l|}{} & & & & & & & z: 26.2, y: 24.9 (5$\sigma$, point source)\\

\tableline
\end{tabular}
\tablenotetext{a}{Data from Euclid Definition Study Report (\cite{2011arXiv1110.3193L}.)}
\tablenotetext{b}{Data from WFIRST-AFTA Final Report (\cite{2013arXiv1305.5422S}).}
\tablenotetext{c}{Data from LSST science book (\cite{2009arXiv0912.0201L}).}
\end{table*}

There are many astronomical image simulation tools. GalSim (\cite{2015A&C....10..121R}) is an open-source package for simulating images of stars and galaxies using a range of methods include Sérsic profile and Shapelets to describe galaxies. The Photon Simulator (PhoSim) (\cite{2015ApJS..218...14P}) is LSST optical image simulator which uses Monte Carlo method to calculate the physics of the atmosphere and a telescope and camera in order to simulate realistic astronomical images. SkyMaker (\cite{2009MmSAI..80..422B}) is a software package for creating artificial astronomical images which also includes the features of an instrument (CCD, optical system, et al.), noise models, the galaxy profile (disk + bulge) and so on. \cite{2010PASP..122..947D} show an image software suite for the simulation extragalactic images based on real observations of HUDF and their pipeline includes the multi-wavelength catalog generation, the repopulation of the catalog images and the addition of the various noise components, but the simulation image of this software will be produced only in the B, V, i, and z bands.
In this paper, for convenience and getting intermediate results of calculations, we develop a new image simulator for CSS-OS which includes the galaxy simulation, the stellar simulation, the instruments, and noises and background features. Moreover, our simulator can also perform data analysis.

The rest of this paper is organized as follows. Section \ref{sec:catalog} introduces HUDF catalogs characteristics and the guideline of selecting galaxy samples for CSS-OS and LSST. Section \ref{sec:mehtod} describes the methods of image simulation, including PSF model, galaxy profile model, and noise features. In Section \ref{sec:result_analysing}, we present the simulation result for CSS-OS and LSST and the analysis of the result. Finally, we conclude and summarize in Section \ref{sec:summary}.

\section{Galaxy Catalog}
\label{sec:catalog}

\subsection{Galaxy catalog overview}
In this paper, the effective galaxy density is evaluated by the analysis from mock galaxy images. The galaxy samples added to the mock image must meet the completeness of the galaxies detection for CSS-OS and LSST. Whether a galaxy can be detected is mainly decided by the surface brightness of the galaxy, the aperture of a telescope and integration time used in observation. The surface brightness is inherent in galaxies, so the integration time is very important in galaxy detection for the telescope. The galaxy catalog used in our image simulation must be extracted from the image of deep field observation which is obtained with long enough integration time for getting enough flux from galaxies so that catalogs include all of the galaxies which can be detected by CSS-OS and LSST. 

The Cosmic Evolution Survey(COSMOS)\footnote{\url{http://cosmos.astro.caltech.edu/}} is a deep, wide area, multi-wavelength survey, whose galaxy catalogs (\cite{2009ApJ...690.1236I}, \cite{2016ApJS..224...24L}) include photometry and photometric redshift. Though cosmos catalogs are very deep and meet the request of CSS-OS, it can't reach the depth of LSST for its short accumulated exposure time $\sim$ 2000s which is less than 1/3 of that of LSST. Another famous deep survey is the Great Observatories Origins Deep Survey (GOODS)\footnote{\url{http://www.stsci.edu/science/goods/}} which covers two fields centered on the Hubble Deep Field North (HDF-N) and the Chandra Deep Field South (CDF-S) and uses a number of facilities include NASA's Spitzer, Hubble, and Chandra, ESA's Herschel and XMM-Newton, and some most powerful ground-based facilities. In GOODS-S area (CDF-S), HST chooses a field of an over 11 arcmin$^2$ region where the observation condition is very good called Hubble Ultra Deep Field (HUDF) (\cite{2006AJ....132.1729B}) to observe faint galaxies. HUDF is maximum overlap with extant or planned deep observations in different bands and the efficiency of observation is very high for HST. There are 4 ACS bands: $B_{435}$ (F435W, hereafter B), $V_{606}$ (F606W, hereafter V), $i_{775}$ (F775W, hereafter HUDF-i), and $z_{850}$ (F850LP, hereafter HUDF-z) and the total exposure time is just under 1 million seconds from 400 orbits. Due to the long integration time and high sensitivity of ACS, most of the signals of the faint objects can be detected from the image. There are over 10,000 objects detected from the HUDF image, most of which are faint objects including the high redshift galaxies (high redshift to about 6) and the low surface brightness galaxies (LSBs). In last years, large numbers of the ultra-diffuse galaxies (UDGs) have been found (\cite{2015ApJ...798L..45V},~\cite{2015ApJ...807L...2K}). The UDGs is one of the LSBs, but the surface brightness of UDGs is lower than that of “classical” LSBs. \cite{2012AJ....143..102G} show that lower surface brightness objects dominate the faint-end slope of the luminosity function in the field and they believe that the surface brightness limits is important in evaluating measurements of the faint-end slope. Long integration time, like HUDF observation, can improve the surface brightness limits for detecting the LSBs especially the UDGs. Therefore, the HUDF image contains most of high redshift galaxies and LSBs. In the latter part, we also demonstrate the completeness of HUDF galaxy samples for CSS-OS and LSST galaxy analysis by comparing  pixel signal-to-noise ratios.

\subsection{The completeness of HUDF catalog for the analysis of CSS-OS and LSST}
MultiDrizzle program (\cite{2002hstd.book.....K}) is used to combine all the images multi-exposure of an object or a sky area for each band to create a clean median and high pixel resolution image. The pixel resolution of the drizzled HUDF image is about 0.03". Table \ref{tab:HUDF_info} gives some information about HUDF include total exposure time and PSF of each band. From the drizzled HUDF-i image, we extract background information with the SExtractor program (\cite{1996A&AS..117..393B}), and the RMS of background is 8.1$\times10^{-4} s^{-1}pixel^{-1}$ which is the noise of HUDF-i drizzled image (including the background noise and the instrumental noise). From Table \ref{tab:sim_param}, we can obtain the simulation parameter of CSS-OS and LSST. The galaxy is an extended source which  contains multiple pixels in the image. If the galaxy can be detected from an image, most of the galaxy image pixels need to be extracted. Assuming the galaxy is uniform, the flux of each galaxy image pixel is the same. Then we use the pixel signal-to-noise ratio (SNR) to test whether the mock image with HUDF catalogs meets the requirement of galaxy detection completeness by CSS-OS and LSST.

\begin{table*}
 \caption{HUDF information}
 
 \label{tab:HUDF_info}
 \begin{tabular}{l|llll}
   \tableline
   & $B_{435}$ & $V_{606}$ & $i_{775}$ &  $z_{850}$\\
   \tableline
  Total orbits & 56 & 56 & 144 & 144 \\
  Total observation time (s) & 134900 & 135300 & 347100 & 346600 \\
  FWHM of PSF(") & 0.084 & 0.079 & 0.081 & 0.089 \\
   \tableline
 \end{tabular}
\end{table*}

\begin{table*}
\small
\caption{Parameters of CSS-OS and LSST in Simulation}
\label{tab:sim_param}
\begin{tabular}{l|ll}
\tableline
& CSS-OS & LSST\\
\tableline
Aperture & 2m & 8.4m (effective aperture 6.7 m) \tablenotemark{a}\\
PSF & 0.15"($R_{80}$) & 0.67"(FWHM) \tablenotemark{b}\\
Pixel size & 0.075" & 0.2" \tablenotemark{a}\\
Dark current  & 0.02 e-/pixel/s & 0 \\
Readout Noise & 5 e-/pixel & 10 e-/pixel \tablenotemark{c}\\
Total exposure time & 150s$\times$2 (Large area survey) & 15s$\times$2$\times$230 \tablenotemark{a}\\
& 250s$\times$8 (Deep survey) &\\
Survey cover area & 17500 deg$^2$ & 18000 deg$^2$\\
\tableline
\end{tabular}
\tablenotetext{a}{Data from LSST science book (\cite{2009arXiv0912.0201L}).}
\tablenotetext{b}{The seeing of LSST is measured at 500 nm based on ten years of measurements from CTIO (10 km from the LSST site).}
\tablenotetext{c}{Data from \url{https://www.lsst.org/about/camera/features}, low readout noise $<$ 10 e-/pixel.}

\end{table*}

The SNR of single pixel can be calculated by 
\begin{equation}
\label{equ:snr1}
SNR_{pix} = \frac{S_{pix}\times t}{\sqrt{S_{pix}\times t + \sigma^2_{noise}}}
\end{equation}
where S$_{pix}$ is the signal from the astronomical source in a single pixel, t is total exposure time, $\sigma_{noise}$ is the noise except for signal shot noise. If the noise consists of signal shot noise, background noise, dark current noise and readout noise, Eq. \ref{equ:snr1} can be written as

\begin{equation}
\label{equ:snr2}
SNR_{pix} = \frac{S_{pix}\times t}{\sqrt{S_{pix}\times t + B_{pix}\times t + D\times t + N_{read}\times R^2}}
\end{equation}
where $B_{pix}$ is the value of the background in a single pixel, $D$ is the value of dark current which also can be ignored for low environmental temperature, $R$ is readout noise and $N_{read}$ is the number of pixels of CCD readout.

\begin{figure}[!htpb]
 \includegraphics[width=\columnwidth]{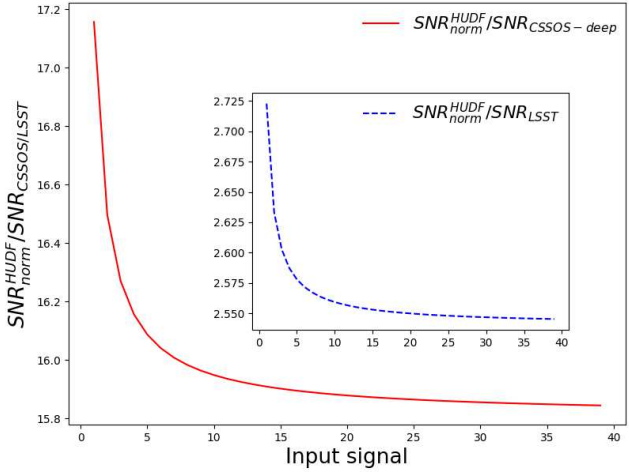}
 \centering
 \caption{Pixel SNR comparison between HUDF and CSS-OS, LSST. Assuming the input signal is from a uniform extended source, so the PSF needn’t  be considered in this calculation. The signal received by the detector can be obtained by aperture, total exposure time and telescope efficiency, then use the pixel size of the different telescope to normalize the received signal. The red solid line represents the ratio between HUDF pixel SNR normalized by CSS-OS pixel size and CSS-OS pixel SNR, and blue dashed line represents the ratio between HUDF pixel SNR normalized by LSST pixel size and LSST pixel SNR.}
 \label{fig:snr_comp}
\end{figure}

Due to the difference in pixel size between the detectors of HUBBLE ACS, LSST, and CSS-OS, we need to normalize pixel SNR of HUDF image to the same pixel scale with CSS-OS or LSST to check whether the galaxy catalog of HUDF fits CSS-OS or LSST in the aspect of galaxy detection completeness. For the SNR estimation, we assume that galaxy is a uniform extended source, so if normalize the pixel SNR of HUDF to the pixel SNR of CSS-OS or LSST, it can be written as

\begin{equation}
\label{equ:area}
SNR^{HUDF}_{norm} = \frac{S^{norm}_{pix}\times t}{\sqrt{S_{pix}^{norm}\times t + \frac{A_{pix}^{norm}}{A_{pix}^{HUDF}}\times \sigma^2_{HUDF}}}
\end{equation}
where S$^{norm}_{pix}$ is normalized signal by CSS-OS pixel size or LSST pixel size, which could be affected by the aperture of the telescope. A$^{norm}_{pix}$ represents the normalized area determined by the pixel size of CSS-OS or LSST, A$_{pix}^{HUDF}$ represents the pixel area of HUDF. Fig. \ref{fig:snr_comp} shows the SNR comparison. The red solid line represents the ratio between HUDF pixel SNR normalized by CSS-OS pixel size and CSS-OS pixel SNR and the blue dashed line represents the ratio between HUDF pixel SNR normalized by LSST pixel size and LSST pixel SNR. From the result, we suppose that SNR of galaxy detected by HUDF is higher than that of CSS-OS and LSST. In our calculation, the integration time of CSS-OS uses the time of deep survey (2000 s). If HUDF catalog can meet SNR request of CSS-OS deep survey, it must meet the request of CSS-OS large area survey (300s) in the simulation. Therefore,  we consider that the HUDF catalogs comprise all galaxies detected by CSS-OS and LSST.

\subsection{Galaxy sample selection}
\cite{2006AJ....132..926C} give a set of catalogs include multi-band photometry catalog, photometric redshifts catalog, and morphology catalog. They use ACS B, V, HUDF-i, HUDF-z, and NIC3 J, H band images and remap the NIC3 segmentation maps to the ACS frame, then combine detections from HUDF-i, HUDF-z, J + H, and B + V + HUDF-i + HUDF-z images into a single comprehensive segmentation map to obtain new catalogs. There are $\sim$18700 objects in photometry catalog and photometric redshifts catalog.

The morphology catalog is extracted from HUDF-i image (the deepest ACS image). And the shape profile of this catalog is described with Sérsic profile which includes Sérsic index, effective radius, position angle, and ellipticity. There are 9069 extracted objects with SExtractor stellarity $<$ 0.9 in the morphological catalog which is also compared with \cite{2006AJ....132.1729B} catalog (hereafter B04, where 2004 refers to the release date) based on ACS image. Removing the objects that mismatch with B04, there are 8805 objects left. There are 124 objects undetected by \cite{2006AJ....132..926C} and 48 objects without HUDF-i band photometry data, so 8633 objects survive in finally. 

\begin{figure}[!htpb]
 \includegraphics[width=\columnwidth]{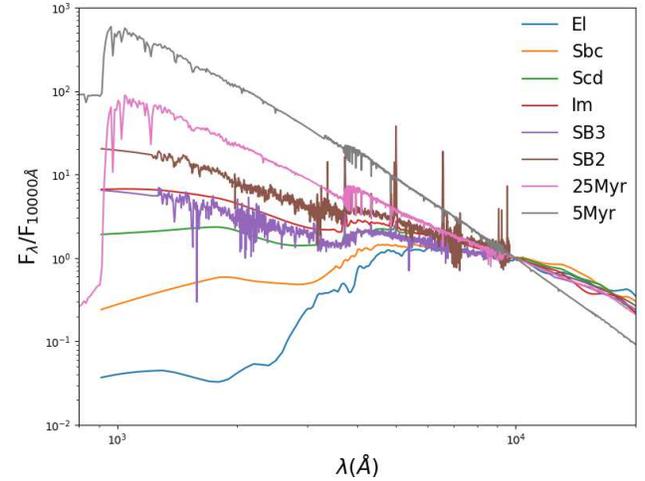}
 \centering
 \caption{SED templates used in this paper which is from \cite{2006AJ....132..926C}. There are eight templates. The templates from El to Sb2 are given in  \cite{2004ApJS..150....1B} and the other two more blue faint galaxies templates, 25 Myr and 5 Myr, are produced by BC03 (\cite{2003MNRAS.344.1000B}). These SEDs consist of galaxies with different SFR histories.}
 \label{fig:seds}
\end{figure}

\subsection{Mock galaxy spectrum}
\begin{figure}[!htpb]
 \includegraphics[width=\columnwidth]{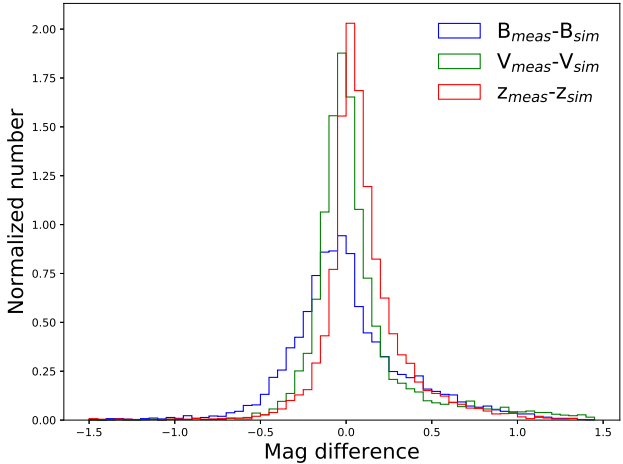}
 \centering
 \caption{The difference distribution between the photometric data and the simulation data from B, V, HUDF-z bands. Using the mock galaxies spectrum with the throughput of B, V, HUDF-z bands in Hubble telescope system, we can calculate the brightness of these bands. }
 \label{fig:mag_comp_BVz}
\end{figure}

\cite{2006AJ....132..926C} have used the photometry data of ACS B, V, HUDF-i, HUDF-z, and NIC3 J, H bands to produce photometric redshifts catalog by BPZ (\cite{2000ApJ...536..571B}). From the photometric redshifts catalog, we can obtain the photometric redshift and SED type for all selected objects. It uses eight SEDs including elliptical galaxies, spiral galaxies, irregular galaxies, starburst galaxies, and faint blue galaxies. Fig. \ref{fig:seds} shows the SED template sets used in this paper. These template sets include templates from the faint blue galaxy (5 Myr SSP SED) to the elliptical galaxy (El) whose star formation histories cover from early to late formation. In this paper, we use the SED templates and photometric redshifts to mimic the galaxy spectrum for each selected galaxy. Then each galaxy spectrum is normalized to proper flux according to its photometric data. All the photometric data of the catalog are in AB magnitude system (\cite{1983ApJ...266..713O}), and the F$_{\upsilon}$ is constant which is in units of ergs cm$^{-2}$ s$^{-1}$ Hz$^{-1}$. The spectrum flux normalization process as

\begin{equation}
\label{equ:norm_ab1}
factor_{norm} = \frac{\int^{\lambda_2}_{\lambda_1} {\frac{F_{AB}(\lambda)10^{-0.4m}T(\lambda)}{E(\lambda)}d\lambda}}{\int^{\lambda_2}_{\lambda_1} {\frac{F_{spec}(\lambda)T(\lambda)}{E(\lambda)}d\lambda}}
\end{equation}
and 

\begin{equation}
\label{equ:norm_ab2}
F(\lambda) = factor_{norm} \times F_{spec}(\lambda)
\end{equation}
where $F(\lambda)$ is the flux of the normalized spectrum, $F_{spec}(\lambda)$ is the flux of original spectrum, $F_{AB}(\lambda)$ is the flux in zero AB magnitude, $T(\lambda)$ is band throughput, $m$ is AB magnitude of the galaxy and $E(\lambda)$ is the energy of photon. $F(\lambda), F_{spec}(\lambda)$ and $F_{AB}(\lambda)$ are in units of ergs cm$^{-2}$ s$^{-1}$ $\lambda^{-1}$ and $F_{AB}(\lambda)$ can be calculated by the constant of F$_\upsilon$. Using the methods above, we can produce the spectrum for each galaxy. Weighting the spectrum with the throughput of the different bands, the brightness of the galaxy can be calculated in the different bands. We compare the photometric data of B, V, HUDF-z with simulation data derived by mocking spectrum of each galaxy. Fig. \ref{fig:mag_comp_BVz} shows the difference distribution between the photometric data and the simulation data from B, V, HUDF-z bands (due to using the photometric data of HUDF-i for the spectrum normalization, the brightness of the HUDF-i  simulation data is the same as the inputting HUDF-i photometric data). The peak of the difference distribution for B, V, HUDF-z are in $\sim$ -0.03, $\sim$ -0.03 and $\sim$ 0.03 AB magnitude respectively. The differences mainly come from the photometric redshift procedure which uses eight templates (Fig. \ref{fig:seds}). The observed color can be anything, while the model color is the one from the closest template. Therefore, there are considerable differences in other bands other than i band. The HUDF-i band and HUDF-z band cover the range of CSS-OS i band and LSST i band, so the errors between the photometric data and the simulation data of CSS-OS i band and LSST i band are less than 0.03 AB magnitude.

\subsection{The selected galaxy attributes}
\begin{figure}[!htpb]
 \includegraphics[width=\columnwidth]{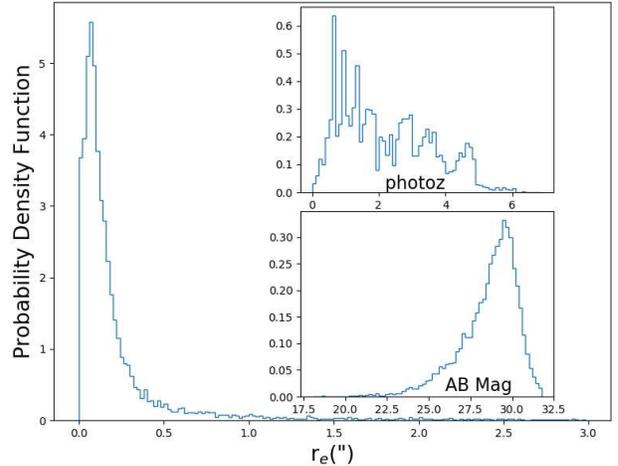}
 \centering
 \caption{Selected galaxies attributes. It shows the histogram of the effective radius(r$_e$), photometric redshift (photoz) and magnitude. The bin widths are 0.02, 0.1 and 0.2 respectively.}
 \label{fig:gal_attr}
\end{figure}

Fig. \ref{fig:gal_attr} shows the selected galaxy attributes. Our selected galaxies are all effective galaxies, from which we can get photometric redshift and shape parameters by measurement. The effective radius is obtained using GalFit (\cite{2002AJ....124..266P}) with Sérsic model by \cite{2006AJ....132..926C}. The peak of r$_e$ histogram is at r$_e$ $\sim$ 0.07". Since r$_e$ has a maximum of 500 pixels (15",1 pixel $\sim$ 0.03") in GalFit simulation, the value of r$_e$ extends to 15" but the ratio of which is quite small. From the redshift distribution of the selected galaxies, we can see that the range of redshift is from 0 to 6 and the peak of redshift appears at z $\sim$ 0.65. The selected galaxies have deep magnitude limit, and the peak of magnitudes distribution is at mag$_{AB}$ $\sim$ 29.5 AB magnitude and the range of galaxy brightness is from 18 to 32 AB magnitude.

\section{The Method of Image Simulation}
\label{sec:mehtod}
One goal of our simulation is to get high fidelity optical astronomical mock images, using which we can obtain the information of each galaxy. Then by statistics on a large number of galaxies information, we can acquire the difference of galaxy samples between CSS-OS and LSST, and we also get the effective galaxy number count of the two projects. Here, we use some simple but very effective models in our image simulation.

\subsection{Point Spread Function}
The Point Spread Function(PSF) is the description of the response of an imaging system to a point source. For the astronomical image, PSF includes the instrument effect (optical system, filter, detector, et al.) and observation conditions (primarily for ground-based telescopes due to the atmosphere turbulence). We use the Moffat profile(\cite{1969A&A.....3..455M}) to account for PSF which is numerically well behaved and well fitting atmospheric turbulence model. The Moffat profile as follows:
\begin{equation}
 I(r) = I_0[1+(\frac{r}{\alpha})^2]^{-\beta}
\end{equation}
where $I_0$ is the flux density at r = 0, $\alpha$ and $\beta$ are scale parameters. The relationship between $\alpha$ and $\beta$ is
\begin{equation}
 \alpha = \frac{FWHM}{2\sqrt{2^{\frac{1}{\beta}}-1}}
\end{equation}
where $FWHM$ is the abbreviation of Full Weight at Half Maximum, so $I(\frac{FWHM}{2})=\frac{1}{2}I_0$. In general, the value of $\beta$ can be used to describe observation conditions. When $\beta = 4.765$(\cite{2001MNRAS.328..977T}) Moffat PSF can fit well with the atmospheric turbulence model.

The space-based telescope has high angular resolution since it is not affected by atmospheric turbulence, so the PSF of the space-based telescope is steeper than that of the ground-based telescope in the condition that the optical system is perfect. For Moffat profile, the profile becomes steeper as $\beta$ becomes larger. \cite{2001MNRAS.328..977T} also show how Moffat PSF contains Gaussian PSF as a limiting case when $\beta \rightarrow \infty$. In general, we use a Gaussian profile to describe the PSF of space telescope (in this paper, for CSS-OS) and a value of $\beta$ = 100 is completely satisfactory for modeling a Gaussian by using the Moffat function.
\subsection{Galaxy Profile}
There are many methods to describe galaxy profile, such as Gaussian or Moffat function(\cite{1969A&A.....3..455M}) , Sérsic profile (\cite{1963BAAA....6...41S}), De Vaucouleurs profile (\cite{1948AnAp...11..247D}) (one case of Sérsic profile), shapelets (\cite{2003MNRAS.338...35R},\cite{2003MNRAS.338...48R}). GalFit (\cite{2002AJ....124..266P}) is a tool for fitting galaxy profile and extract structural components from an image which includes galaxy profile models above except for shapelets. Shapelets is a complicated method of galaxy image decomposition, and it needs to decompose the object in the image into a series of localized basis functions of different shapes.  In this paper, the galaxy shape components are extracted by GalFit using Sérsic profile. Therefore, we use Sérsic profile to rebuilt galaxy images in our simulation. The Sérsic profile is as follow:
\begin{equation}
 I(r)=I_ee^{-b_n[{(\frac{r}{r_e})}^\frac{1}{n}-1]}
  \label{equ:sersic1}
\end{equation}
where $r_e$ is the effective radius and $I_e$ is the flux density at $r_e$. $n$ is the Sérsic index which can control the dispersion of galaxy flux. $b_n$ is a constant. Eq. \ref{equ:sersic1} can be transformed as:
\begin{equation}
 I(r)=I_0e^{-{(\frac{r}{a})}^\frac{1}{n}}
  \label{equ:sersic2}
\end{equation}
where $I_0$ is the flux density at r = 0, $a$ is the radius at which flux density drops as  $e^{-1}$ of $I_0$.

In general, we use elliptic shape to denote galaxy shape. We set $e$ as ellipticity, and the normalized elliptic profile can be described as:
\begin{equation}
r^2 = \frac{x^2}{1-e}+(1-e)y^2
\label{equ:transR}
\end{equation}
In there, $[x, y]$ is image coordinate. If r is fixed, we can obtain the isoline in which the surface brightness is the same.

If we use the image coordinate to describe a galaxy profile, there exists a position angle between galaxy coordinate and image coordinate. Then we need to use the transformation matrix

\begin{equation}
\left( \begin{array}{cc}
    \cos(\theta)&-\sin(\theta) \\
    \sin(\theta)&~~\cos(\theta) 
\end{array} \right) 
\end{equation}
to transform the galaxy coordinate to image coordinate, where $\theta$ is the position angle. Then we can get the new galaxy coordinate $[x^{'},y^{'}]$ in an image as:
\begin{equation}
\left( \begin{array}{c}
    x^{'}\\
    y^{'}
\end{array} \right)
=
\left( \begin{array}{cc}
    \cos(\theta)       & -\sin(\theta) \\
    \sin(\theta)       & ~~\cos(\theta)
\end{array}  \right)
\left( \begin{array}{c}
    x\\
    y
\end{array}  \right)
\label{equ:transNewCoor}
\end{equation}

Through the Eq. \ref{equ:sersic1} to \ref{equ:transNewCoor}, we can describe the galaxy profile (include flux density distribution, shape and position angle) by using the image coordinate.

\subsection{Noise} 
Except for the PSF model and the galaxy profile model, there is also another significant model for astronomical image simulation that is the noise model. In our simulation, there are two kinds of noises: one is photon shot noise, another is instrumental noise. Photon shot noise mainly arises from astronomical object signal and sky background light. Photon shot noise is a spatially and temporally random phenomenon described
by Bose-Einstein statistics (\cite{2007ptd..book.....J}) as:
\begin{equation}
\sigma_{phot}^2=F_{phot}\frac{e^{\frac{E}{kT}}}{e^{\frac{E}{kT}}-1}
\end{equation}
where $F_{phot}$ is interacting photon, $E$ is the energy of a photon, $k$ is Boltzmann’s constant (1.38 $\times$ 10$^{-23}$ J/K), and $T$ is absolute temperature (K). Assuming $E\gg kT$, then 
\begin{equation}
\sigma_{phot}=\sqrt{F_{phot}}
\end{equation}
While, for the instrumental noise, we add dark current noise and readout noise. Dark current is relatively small electric current accumulated over time. Dark current noise follows a Poisson distribution, the RMS of dark current noise is the square root of the dark current. Readout noise is only related to the readout circuit regardless of exposure time. We assume that the readout noise distribution corresponds to a Gaussian distribution. Besides, there exists thermal noise for an instrument, but it can be cut down by lowering the temperature of the chip, so we ignore the effect of thermal noise in our calculation.

\section{Simulation Result and Data Analyzing}
\label{sec:result_analysing}

\subsection{The simulation conditions and results} 
In Section \ref{sec:catalog}, we produce the catalog used in this paper which contains morphological information, magnitude and spectrum of each galaxy selected. In Table \ref{tab:sim_param}, we give the parameters used in the image simulation for CSS-OS and LSST. Fig. \ref{fig:fil_throughput_magcheck} shows the i band throughput of CSS-OS and LSST. For CSS-OS, it contains the optical system efficiency, quantum efficiency of CCD, and i band filter efficiency. While for LSST \footnote{\url{https://github.com/lsst/throughputs/tree/master/baseline}}, the atmospheric extinction in its i band throughput also needs to be considered. In our simulation, we also need to take into account the sky background. CSS-OS is a space-based project, so the sky background mainly consists of the zodiacal light and the earthshine light mostly. We estimate the sky background of CSS-OS using the data of Hubble space telescope (\cite{2012acsi.book.....U}) which provides the curve of zodiacal light and earthshine light. In this paper, we use the average sky background of Hubble. While LSST is a ground-based project, the sky background is affected by the target - moon angular separation, lunar phase, ecliptic latitude, zenith angle, and phase of the solar cycle. A model optical sky spectrum we used is scaled by the broad-band sky brightness from Gemini observatory\footnote{\url{https://www.gemini.edu/sciops/telescopes-and-sites/observing-condition-constraints/optical-sky-background}}. We use the 'darkest' background fainter than which for 20$\%$ of the time for any random target.

\begin{figure*}[!htpb]
 \includegraphics[width=\textwidth]{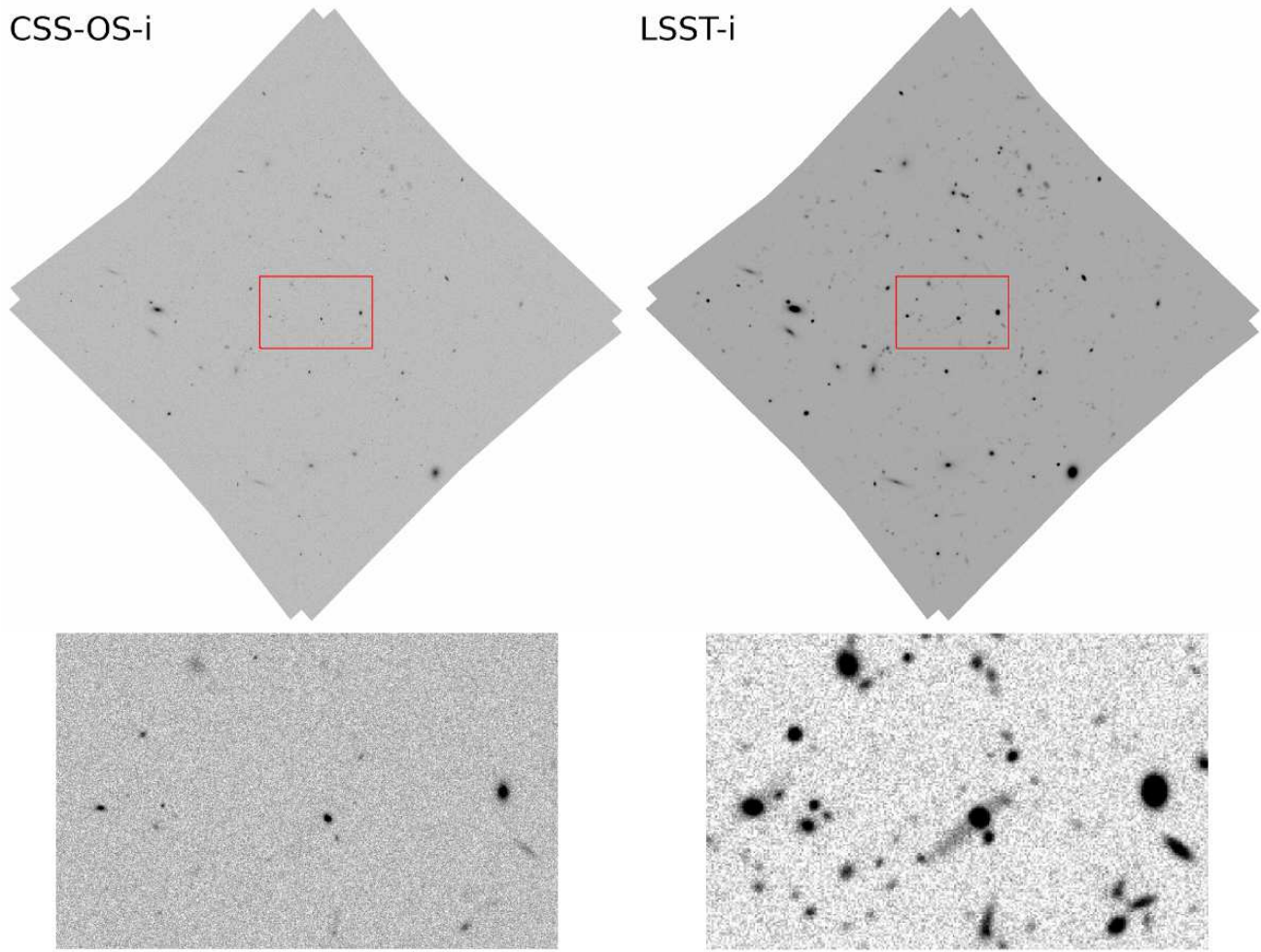}
 \centering
 \caption{Mock images with HUDF catalog. Left: CSS-OS i band mock images. Right: LSST i band mock images. Top: The mock images produced by HUDF catalog whose area are 11.97 arcmin$^2$. Bottom:  The enlarged views for the field of red box in top image whose area are about 0.5 arcmin$^2$.}
 \label{fig:image_sim}
\end{figure*}

Fig. \ref{fig:image_sim} shows the simulation images. The area of the images is about 11.97 arcmin$^2$. The top-left is the simulation image of CSS-OS i band (for large area survey), and the top-right is the simulation image of LSST i band. The bottom-left and the bottom-right are the enlarged view of the central part of the simulation image whose area is about 0.5 arcmin$^2$. From the scaled images, we can find that the source density of LSST is larger than that of CSS-OS in visual, and any source looks smaller when detected with CSS-OS with respect to LSST. 

\subsection{The statistical characteristics of detected galaxies for CSS-OS large area survey and LSST} 

We use the SExtractor program (\cite{1996A&AS..117..393B}) to extract sources from simulation images, then match the extracted source with HUDF catalog. Finally, we obtain the source from CSS-OS simulation image or LSST simulation image. If the SNR$_{obj}$ $\ge$ 10 and FWHM$_{obj}$ $\ge$ 1.25 FWHM$_{psf}$ (\cite{2011arXiv1110.3193L}) for the object detected, we consider it as a galaxy, or the object will be discarded.  According to the criteria above, we can calculate the detected galaxy number for CSS-OS and LSST.

\begin{figure}[!htpb]
 \includegraphics[width=\columnwidth]{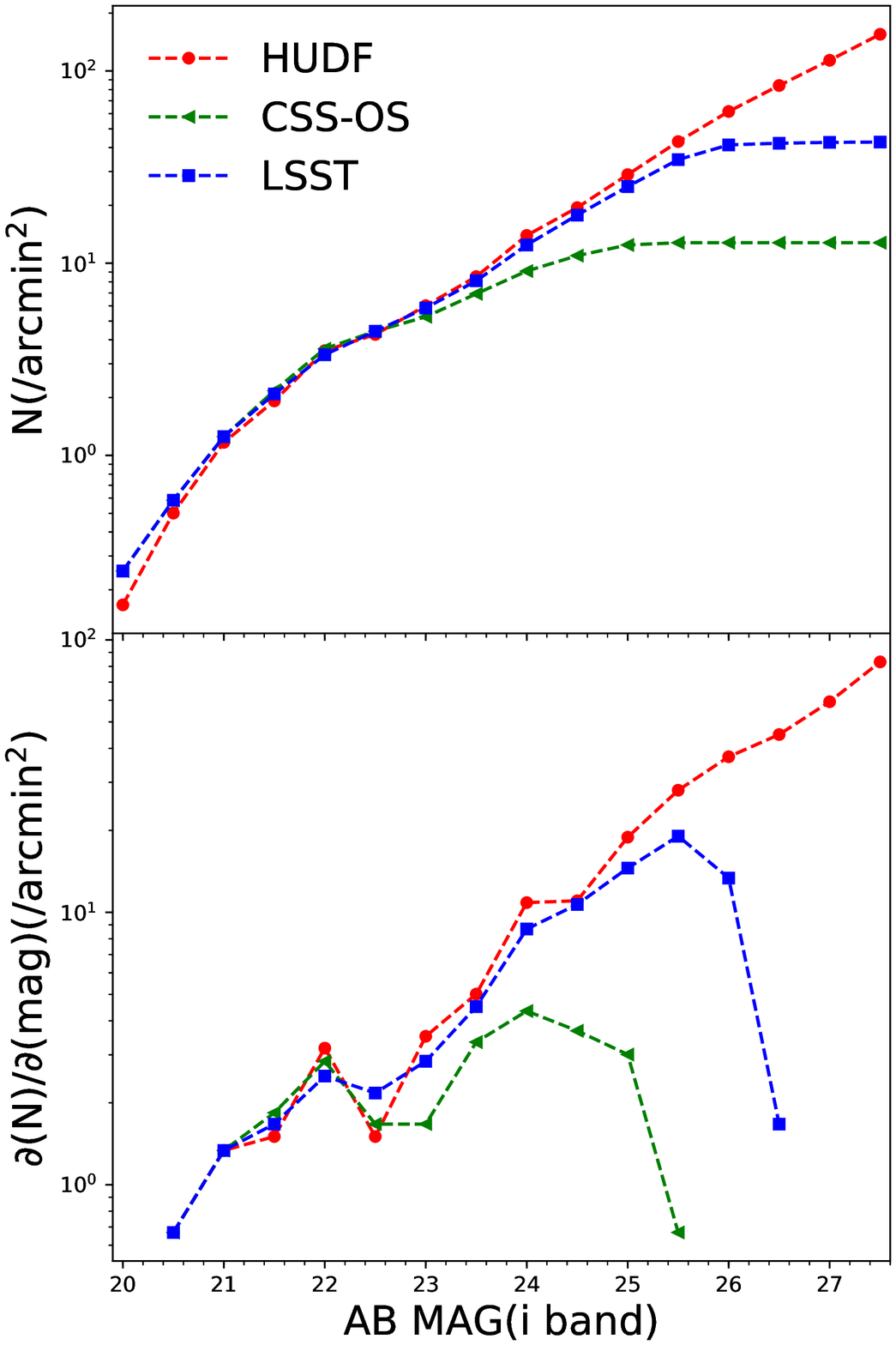}
 \centering
 \caption{The relationship between the effective galaxy number density and the magnitude for detected galaxies. The top is the accumulated number density ($\leq$ magnitude), and the bottom is the differential number density by $\Delta$Mag.}
 \label{fig:gal_num}
\end{figure}

Fig. \ref{fig:gal_num} shows the relationship between the effective galaxies number density and the magnitude for detected galaxies. The red circle points show the galaxies whose SNR $\ge$ 10 from HUDF catalog we selected and there are 5684 galaxies (we only show the data to 27.5 AB magnitude in this figure). The green triangle points are the CSS-OS extracted data, and the blue square points are the LSST extracted data. There is a total of 153 and 511 detected galaxies for CSS-OS and LSST respectively. The top is the accumulated number density ($\leq$ magnitude), from this figure, we can see that the data of HUDF have a little deviation from that of CSS-OS and that of LSST in the bright galaxy detection, due to the difference of band cover range of those devices. While the bright galaxy detection of CSS-OS and LSST is similar. The capacity of LSST galaxy detection is larger than CSS-OS, we use the HUDF-i band data to estimate that the effective number of detected galaxies is $\sim$ 42 arcmin$^{-2}$ for LSST and $\sim$ 13 arcmin$^{-2}$ for CSS-OS. The bottom is the differential number density by $\Delta$magnitude, from this figure, we can conclude that the detected galaxies number count increase almost linearly with the magnitude and this trend changes around $\sim$ 24 AB magnitude for CSS-OS and $\sim$ 25.6 AB magnitude for LSST.

\begin{figure}[!htpb]
 \includegraphics[width=\columnwidth]{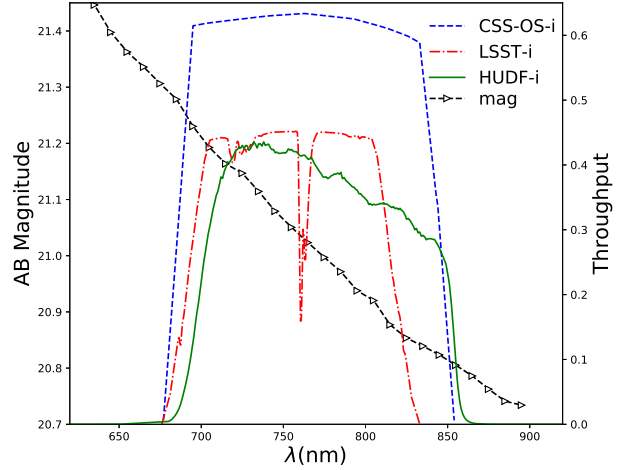}
 \centering
 \caption{The blue dashed line, the red dash-doted line, and the green solid line represents the i band throughputs of CSS-OS, LSST and HUDF respectively. From the selected galaxies, we choose 351 galaxies whose mag$_i <$ 25 for testing and produce the mock spectrum for each galaxy in the test set. All these galaxies spectrums are normalized to 21 AB magnitude in HUDF i band. We use 10 nm as the unit to divide the spectrum of each galaxy to 30 groups from 600 nm to 900 nm and measure the brightness of every unit of 10 nm spectrum part.  By averaging the measured result of each galaxy in every unit of 10 nm spectrum part, we can obtain the black triangle dashed line.}
 \label{fig:fil_throughput_magcheck}
\end{figure}

Fig.~\ref{fig:fil_throughput_magcheck} shows the i band throughputs of CSS-OS, LSST, and HUDF respectively. We can see that the band coverage of the three projects is consistent. However, the band throughput  of HUDF-i is lower than that of other projects in the blue end. From the selected galaxies, we choose 351 galaxies whose mag$_i <$ 25 for testing and produce the mock spectrum for each galaxy in the test set. All these galaxies spectrums are normalized to 21 AB magnitude in HUDF i band. We use 10 nm as the unit to divide the spectrum of each galaxy to 30 groups from 600 nm to 900 nm and measure the brightness of every unit of 10 nm spectrum part.  By averaging the measured result of each galaxy in every unit of 10 nm spectrum part, we can obtain the black triangle dashed line curve in Fig.~\ref{fig:fil_throughput_magcheck}. From this figure, we can find that the red end is brighter than the blue end for the galaxy samples we selected, while we also can see that the efficiency of HUDF-i throughput is lower than that of LSST and CSS-OS in blue end. Therefore, the number density deviation between CSS-OS, LSST, and HUDF for the bright galaxy detection in Fig.~\ref{fig:gal_num} top panel exists.

\begin{figure}[!htpb]
 \includegraphics[width=\columnwidth]{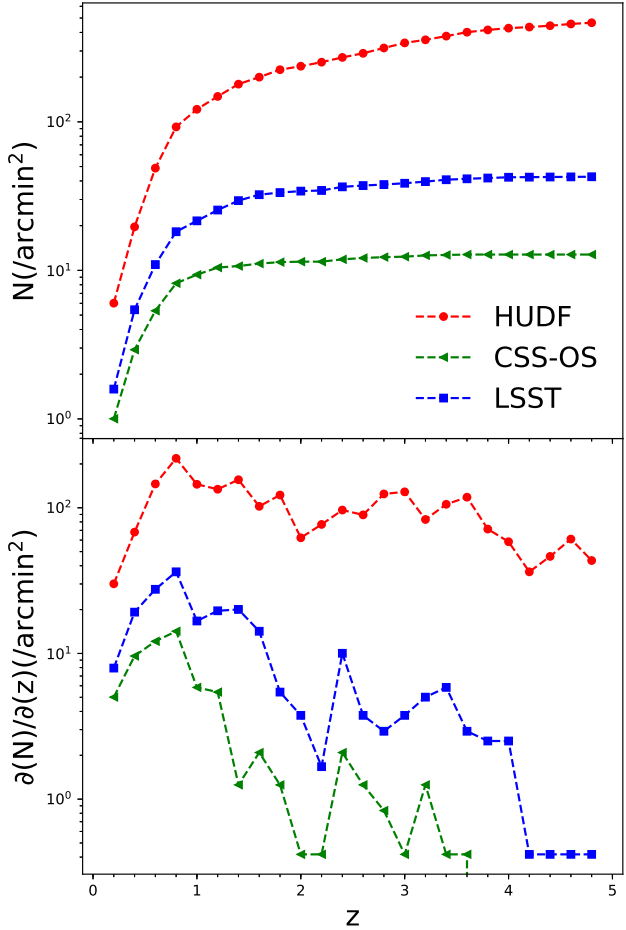}
 \centering
 
 \caption{The relationship between the effective galaxy number density and redshift for detected galaxies. The top is the accumulated number density ($\leq~z$), and the bottom is the differential number density by $\Delta z$. }
 \label{fig:gal_num_z}
\end{figure}

Fig. \ref{fig:gal_num_z} shows  the relationship between the effective galaxy number density and redshift for detected galaxies. From these charts, we can see that the number of detected galaxies increase slowly over z $\sim$ 0.8 and the differential number density of detected galaxies related with redshift drops sharply after z $>$ 0.8. The character of CSS-OS is almost the same as that of LSST. But as redshift grows, the detected galaxies number density between CSS-OS and LSST is larger and larger. Over 60$\%$ of the CSS-OS detected galaxies are at z $<$ 0.8, but only 40$\%$ of the LSST detected galaxies are at z $<$ 0.8. LSST can find higher redshift galaxies while CSS-OS data will powerful from now to z $\sim$ 0.8.

\begin{figure}[!htpb]
 \includegraphics[width=\columnwidth]{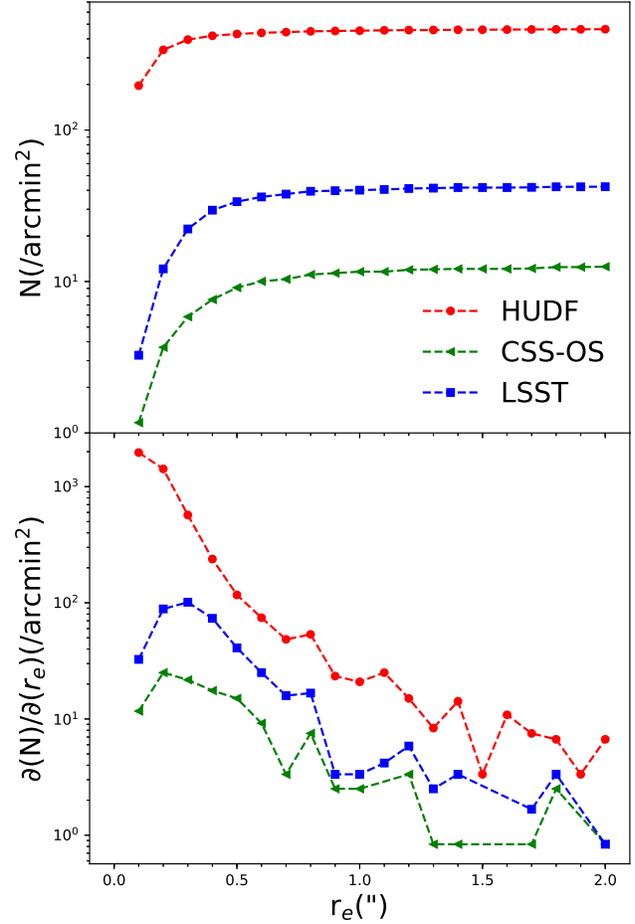}
 \centering
 \caption{The relationship between the effective galaxy number density and the effective radius for detected galaxies. The top is the accumulated number density ($\leq~r_e$), and the bottom is the differential number density by $\Delta r_e$.}
 \label{fig:gal_num_re}
\end{figure}

Fig. \ref{fig:gal_num_re} shows the change of galaxy number density with the effective radius for CSS-OS and LSST. The top shows the relationship between the accumulated galaxy number density and the effective radius. From the top figure, we can find that the detection capability of CSS- OS is very close to that of LSST for small size galaxy detection. As the size of galaxies becomes bigger and bigger, LSST can detect more and more galaxies than CSS-OS dose with a detection limit. The bottom shows the relationship between the differential galaxy number density by $\Delta r_e$ and the effective radius. From the bottom figure, we can see that the peak is at r$_e$ $\sim$ 0.2" for CSS-OS and at r$_e$ $\sim$ 0.3" for LSST, which shows that the sizes of galaxies detected by LSST are larger than that of CSS-OS.  Due to the limit of integration time, CSS-OS can get far fewer galaxies than LSST, but for the detection of small size galaxy detection, the capacity of CSS-OS is not bad.

\begin{figure}[!htpb]
 \includegraphics[width=\columnwidth]{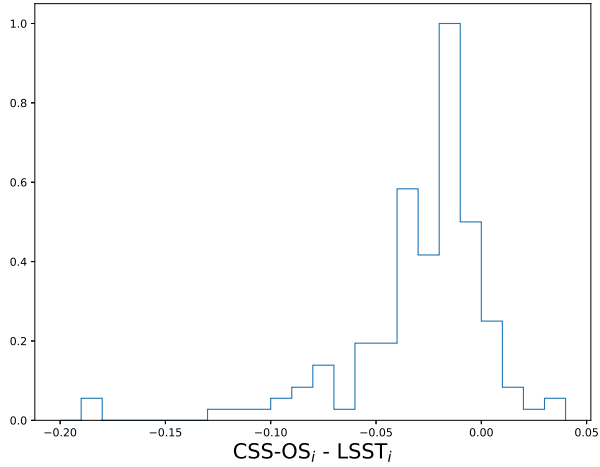}
 \centering
 \caption{The difference between CSS-OS i band photometric system and LSST photometric i band system by the common 135 objects. For the same galaxies, the photometric data of CSS-OS i band is slightly brighter than that of LSST.}
 \label{fig:mag_cssos_lsst}
\end{figure}

We find the common galaxies detected by CSS-OS and LSST and there is a total of 135 common objects for the 511 galaxies detected by LSST and the 153 galaxies detected by CSS-OS. Fig. \ref{fig:mag_cssos_lsst} shows the difference between CSS-OS i band photometric system and LSST i band photometric system by using the common 135 objects. For the same galaxies, the photometric data by CSS-OS i band is slightly brighter than LSST's data. The peak of photometric data deviation between CSS-OS and LSST is $\sim$ -0.015 AB magnitude. We consider that the difference in filters results in the photometric difference of the two projects and we use some typical spectrums to test the photometric difference between i band of CSS-OS and LSST. Table~\ref{tab:i_band_typicalspec} shows the test result. We use four typical spectrums: the elliptical galaxy spectrum in Fig. \ref{fig:seds}, spectrum F$\nu$=constant, vega spectrum with Teff = 9550, log(g) = 3.95, log(z) = -0.5 (\cite{1994A&A...281..817C}) and black-body spectrum with T = 5000k, and normalize the flux of input spectrum to 21 AB magnitude in HUDF-i band. We can find that the photometric results of LSST i band are slightly fainter than that of CSS-OS. In Fig.~\ref{fig:fil_throughput_magcheck}, we choose some galaxies and count the brightness in the different wavelength range. From this result, we can see that the photometric data from CSS-OS i band is slightly brighter than that from LSST due to its larger range in the red end.

\begin{table*}[!htpb]
\caption{Photometric difference between CSS-OS i band and LSST i Band using 4 typical spectrums\tablenotemark{a}}
\label{tab:i_band_typicalspec}
\begin{tabular}{l|llll}
\tableline
& El\tablenotemark{b} & F$\nu$=constant\tablenotemark{c} & Vega\tablenotemark{d} &  Black-body\tablenotemark{e}\\
\tableline
CSS-OS & 21.011 & 20.999 & 20.989 & 21.008 \\
LSST & 21.030 & 21.000 & 20.975 & 21.018 \\
\tableline

\end{tabular}
\tablenotetext{a}{The flux of input spectrums are normalized  to 21 AB magnitude in HUDF-i band.}
\tablenotetext{b}{The elliptical galaxy spectrum shown in Fig. \ref{fig:seds}.}
\tablenotetext{c}{F$\nu$ is spectral flux densities per unit frequency.}
\tablenotetext{d}{Teff = 9550, log(g) = 3.95, log(z) = -0.5, the data from \cite{1994A&A...281..817C}.}
\tablenotetext{e}{The temperature parameter T = 5000k.}
\end{table*}

\begin{figure}[!htpb]
 \includegraphics[width=\columnwidth]{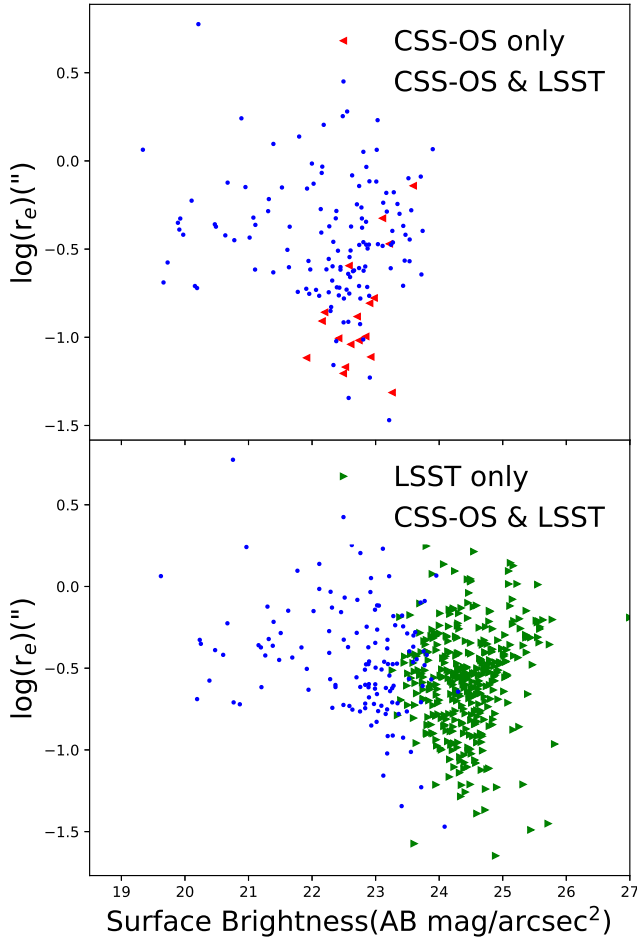}
 \centering
 \caption{The scatter plots of the galaxy surface brightness and the effective radius. The surface brightness is defined as the average surface brightness within the radius of 0.2" (the peak of r$_e$ distribution of CSS-OS) for CSS-OS and the radius of 0.3" (the peak of r$_e$ distribution of LSST) for LSST. The top shows the result extracted from the CSS-OS simulation image and the bottom shows the result extracted from the LSST simulation image. The blue circle points are the galaxies detected by both LSST and CSS-OS, the red triangle points denote the galaxies detected only by CSS-OS, and the green triangle points represent the galaxies detected only by LSST.}
 \label{fig:cssos_lsst_mag_re}
\end{figure}

\begin{figure}[!htpb]
 \includegraphics[width=\columnwidth]{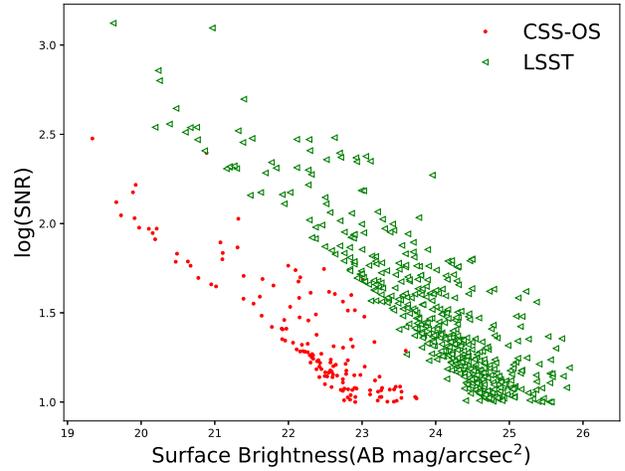}
 \centering
 \caption{The relationship between the galaxy SNR and the surface brightness. The galaxy SNR is calculated from the SExtractor output parameters "FLUX$\_$ISO" and "FLUXERR$\_$ISO". The surface brightness is defined as the average surface brightness within the radius of 0.2" (the peak of r$_e$ distribution of CSS-OS) for CSS-OS and the radius of 0.3" (the peak of r$_e$ distribution of LSST) for LSST.}
 \label{fig:cssos_lsst_snr_sb}
\end{figure}

For the 153 galaxies detected by CSS-OS, there are 18 galaxies un-detected by LSST. While for the 511 galaxies detected by LSST, there are 376 galaxies un-detected by CSS-OS. Fig. \ref{fig:cssos_lsst_mag_re} shows the scatter plots of the galaxy surface brightness and the effective radius. Top shows the result extracted from the CSS-OS simulation image and the bottom shows the result derived from the LSST simulation image. The blue circle points are the galaxies detected by both LSST and CSS-OS, the red triangle points denote the galaxies detected only by CSS-OS, and the green triangle points represent the galaxies detected only by LSST. For the surface brightness calculation of the source extracted from CSS-OS simulation image, we use 0.2" (the peak of r$_e$ distribution of CSS-OS) as the calculation range (top figure). While, the calculation range is within 0.3" (the peak of r$_e$ distribution of LSST), for the sources extracted from the LSST simulation image (bottom figure). From the top figure, we can find that the galaxies detected only by CSS-OS are mostly small galaxies, only 3 out of 18 galaxies have slightly bigger size ($r_e >$ 0.3") which are too close to other galaxies to be distinguished by LSST for its low spatial resolution. From the bottom figure, we can see that there is relatively low surface brightness for the galaxies detected by LSST but CSS-OS not (the green triangle points). CSS-OS has little exposure time so that it is difficult to detect low surface brightness galaxies. Fig. \ref{fig:cssos_lsst_snr_sb} shows the surface brightness as a function of SNR for each galaxy detected by CSS-OS and LSST respectively. The galaxy SNR is calculated from the SExtractor output parameters "FLUX$\_$ISO" and "FLUXERR$\_$ISO". From this figure, the galaxy SNRs from CSS-OS are lower than that from LSST. Therefore, the outer parts of the galaxy image are lost in the noise for the CSS-OS image which also results in that the CSS-OS galaxy images look smaller than LSST. 

\begin{figure}[!htpb]
 \includegraphics[width=\columnwidth]{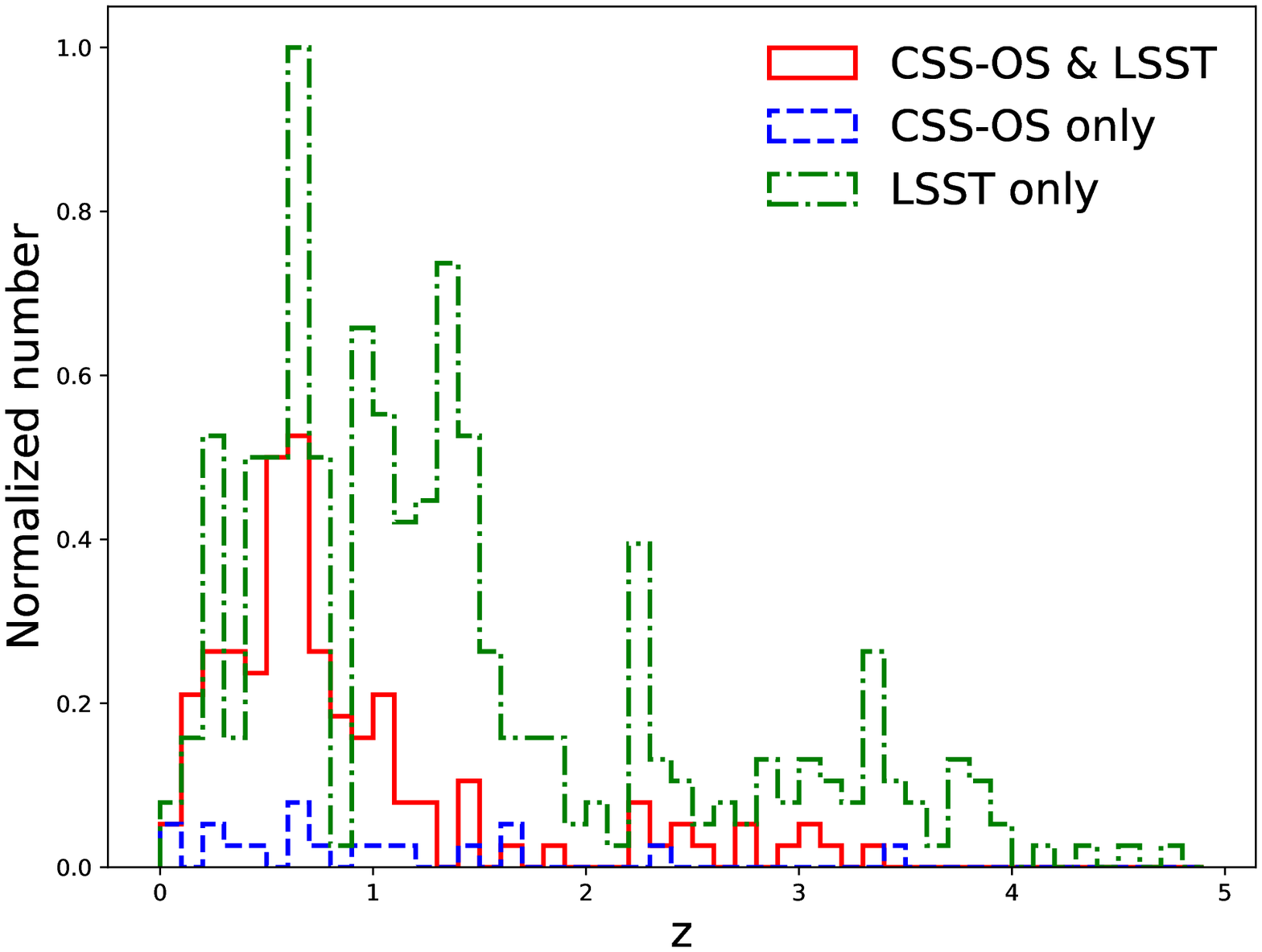}
 \centering
 \caption{The redshift distribution of the source sets detected by CSS-OS and LSST, by CSS-OS only and by LSST only respectively. The red solid line is the distribution of the galaxies detected by CSS-OS and LSST, the blue dashed line  shows the distribution of the galaxies detected only by CSS-OS, and the green dash-dotted line denotes the distribution of the galaxies detected only by LSST.}
 \label{fig:z_cssos_lsst_dif}
\end{figure}

Fig. \ref{fig:z_cssos_lsst_dif} shows the redshift distribution for the source sets detected by CSS-OS and LSST, by CSS-OS only and by LSST only respectively. The redshift distribution of galaxies by CSS-OS only is meaningless because of the small sample size. For the galaxy samples detected by both CSS-OS and LSST, the distribution of the redshift mainly concentrates on the range z $<$ 1. While for the galaxy samples detected by LSST only, the distribution of the redshift distributes in different redshift space that ranges from 0 to 5 and there exits more galaxies in the range z $>$ 1. Due to the stronger observation capabilities, LSST can find farther  and fainter galaxies than CSS-OS. We use the morphological parameter Sérsic index $n$ to distinguish galaxy type. When $n >$ 2.5, they are spheroidal-dominated galaxies, and we think they are early-type galaxies. When $n <$ 2.5, they are disk-dominated galaxies which are considered late-type galaxies. In the galaxy samples detected by CSS-OS, the ratio of early-type galaxies reach $\sim$ 18$\%$, while the ratio is $\sim$ 13$\%$ for LSST. Moreover, In the galaxy samples detected only by LSST, only about 12$\%$ are considered early-type galaxies. Therefore, we think LSST can find more late type galaxies than CSS-OS. 

\subsection{The statistical characteristics of detected galaxies for CSS-OS deep survey} 

\begin{figure}[!htpb]
 \includegraphics[width=\columnwidth]{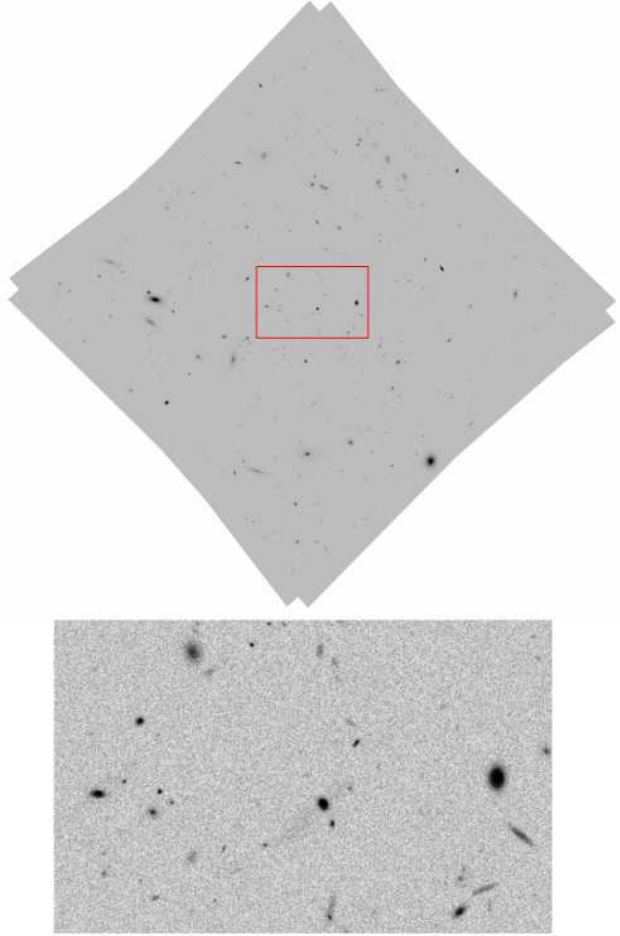}
 \centering
 \caption{The CSS-OS i band deep survey mock image with HUDF catalog. The top image covers all HUDF area about 11.97 arcmin$^2$. The bottom image is the enlarged view for the field of the red box in the top image about 0.5 arcmin$^2$ which is the same as shown in Fig.~\ref{fig:image_sim}.}
 \label{fig:image_cssos_deep}
\end{figure}

Comparing between CSS-OS and LSST, we can find that, for the large area survey of CSS-OS, the capacity of galaxy detection of CSS-OS is much lower than that of LSST. CSS-OS also have deep survey plan for some important areas. For CSS-OS deep survey, the total exposure time reaches 2000 s and the limit detection depth for 5$\sigma$ point source exceeds 27 AB magnitude which is similar to that of LSST.  To discuss the sample difference between CSS-OS as space-based telescope and LSST as ground-based telescope more fairly, we compare the result of CSS-OS deep survey with that of LSST. Using the methods above, we get the CSS-OS deep survey image as shown in Fig.~\ref{fig:image_cssos_deep} and extract galaxies from this image. Then we obtain that the effective galaxy number count is $\sim$ 40 arcmin$^{-2}$ which is close to that of LSST. 

\begin{figure}[!htpb]
 \includegraphics[width=\columnwidth]{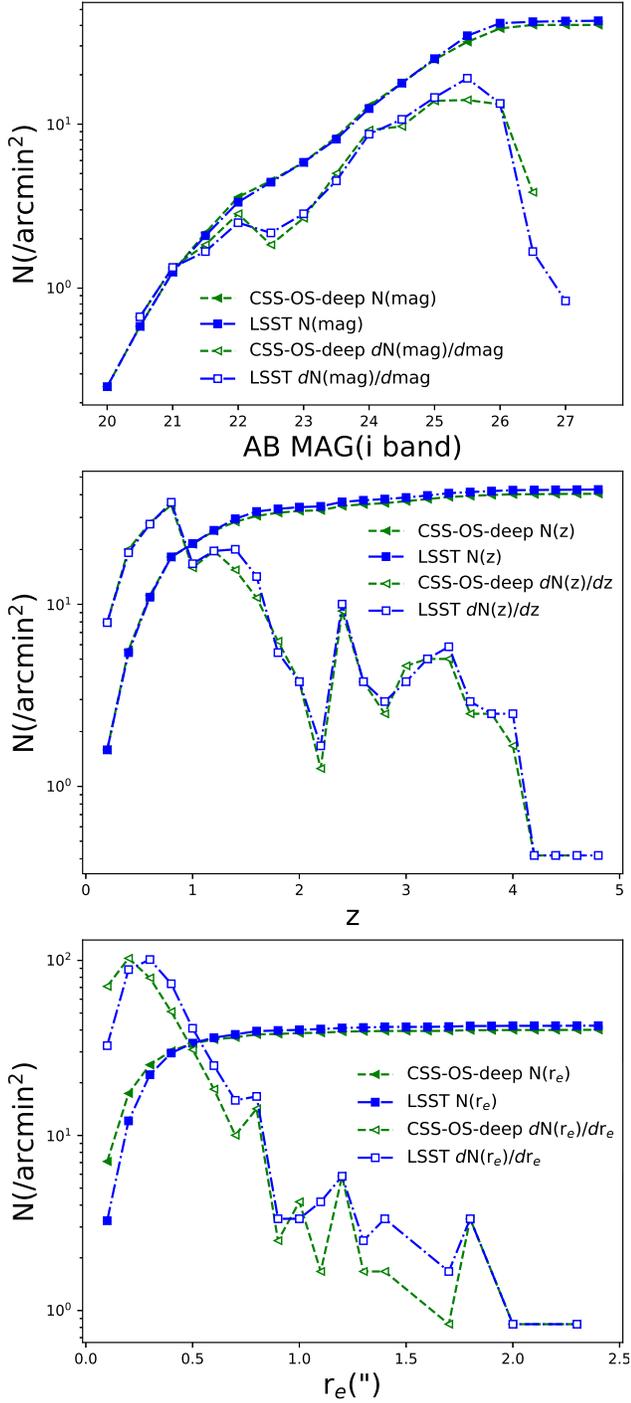}
 \centering
 \caption{The comparison results between CSS-OS deep survey and LSST in terms of i band brightness, redshift and effective radius from top to bottom respectively. The solid point denotes the integration number, and the hollow point denotes the differential number.}
 \label{fig:cssos_deep_mag_z_re}
\end{figure}

Fig. \ref{fig:cssos_deep_mag_z_re} shows the comparison results between CSS-OS deep survey and LSST in terms of i band brightness, redshift and effective radius from top to bottom respectively. The solid points show the accumulated number density, and the hollow points show the differential number density by $\Delta mag$, $\Delta z$ and $\Delta r_e$ respectively. From these figures, we can find that the capacity of galaxies detection of CSS-OS deep survey is close to that of LSST. The capacity of faint galaxies detection of LSST is stronger than that of CSS-OS deep survey for its long integration time from the top figure. But for the small galaxies detection, CSS-OS deep survey as a space-based project has advantages than LSST for its high spatial resolution from the bottom figure. In the redshift space, the redshift distribution of CSS-OS deep survey is similar to that of LSST and the redshift ranges from 0 to $\sim$ 5 for both of them from the center figure. 

\begin{figure}[!htpb]
 \includegraphics[width=\columnwidth]{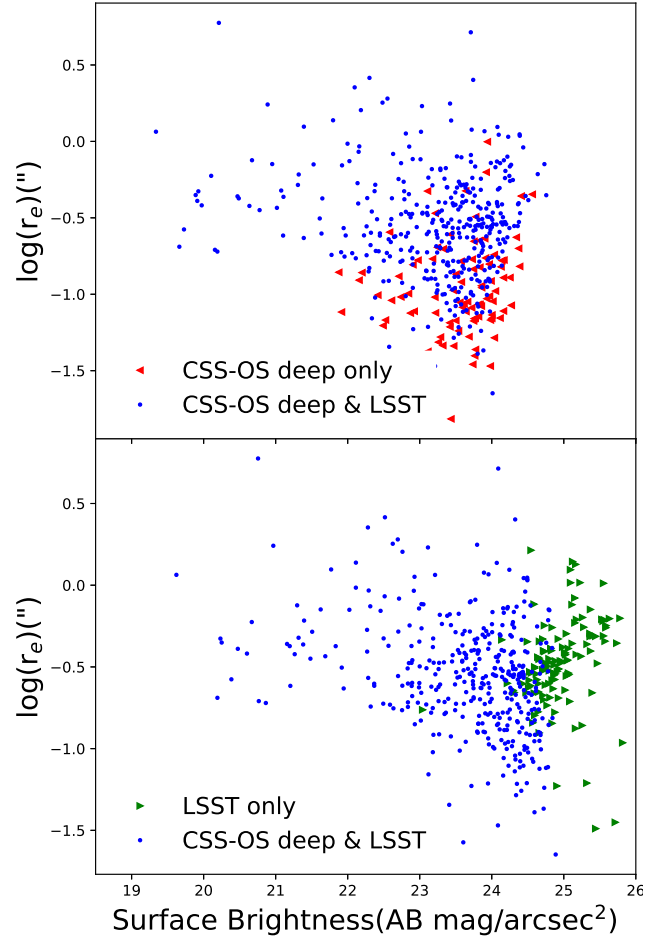}
 \centering
 \caption{The relationship between galaxy surface brightness and r$_e$ for the galaxy samples detected by both CSS-OS deep survey and LSST, only by CSS-OS deep survey and only by LSST respectively. The surface brightness is defined as the average surface brightness within the radius of 0.2" (the peak of r$_e$ distribution of CSS-OS deep survey) for CSS-OS and the radius of 0.3" (the peak of r$_e$ distribution of LSST) for LSST.}
 \label{fig:cssos_deep_mag_re}
\end{figure}

Fig. \ref{fig:cssos_deep_mag_re} shows the relationship between the galaxy surface brightness and the effective radius for the detected galaxies by both CSS-OS deep survey and LSST, only by CSS-OS deep survey and only by LSST. The ratio of common detected galaxies is $\sim$ 80$\%$ and $\sim$ 76$\%$ for the detected galaxies by CSS-OS deep survey and LSST respectively. From Fig. \ref{fig:cssos_deep_mag_re}, we can find that the depth (for a point source) is similar between CSS-OS deep survey and LSST, but LSST is deeper than CSS-OS deep survey for the galaxies detection. There are over 93$\%$ of the galaxies detected only by LSST, whose surface brightness is lower than 24.5 AB magnitude arcsec$^{-2}$. While, for the galaxy samples detected by CSS-OS, there are $\sim$ 99$\%$ of galaxies whose surface brightness is larger than 24.5 AB magnitude arcsec$^{-2}$. LSST uses a larger aperture telescope and a longer exposure time, so it has the advantage for the detection of fainter galaxies. The flux of small galaxies concentrates in the center similar to point source, and CSS-OS has a high spatial resolution so that these small galaxies can be easily distinguished. There are over 90$\%$ of the galaxies detected only by CSS-OS deep survey whose effective radius is smaller than 0.3".

\section{Summary}
\label{sec:summary}
In this work, the HUDF galaxies data are used to simulate the CSS-OS i band data and LSST i band data. We select the HUDF galaxies which can be measured for the shape and the redshift used in the simulation. The data of HUDF from Hubble space telescope and the accumulated exposure time reaches over 340,000 seconds. By the calculation in Fig. \ref{fig:snr_comp}, we think that the HUDF data have enough exposure time to satisfy the completeness of galaxy detection for CSS-OS and LSST. Using the photometric redshift and SEDs, we can produce a mock spectrum for every selected galaxy. Then the galaxy images can be simulated by the spectrum and morphological information combined with the telescope features and the environmental conditions. CSS-OS is a space-based project and LSST is a ground-based project. From the simulation image (Fig. \ref{fig:image_sim} and Fig. \ref{fig:image_cssos_deep}), we can see that the images of CSS-OS and LSST are very different. The space-based telescope has high angular resolution and low noise so that it is very powerful for astronomical observations. The observation is affected by the atmospheric disturbances for the ground-based telescope whose angular resolution is lower than that of  the space-based telescope. 

Using the parameters of Table \ref{tab:sim_param}, we obtain the simulation image for CSS-OS and LSST. The galaxies are extracted by SExtractor, and compared with HUDF catalog for rejecting the detected spurious sources. From the analysis result, we conclude that CSS-OS can detect $\sim$ 13 galaxies arcmin$^{-2}$ which can be measured for shape and redshift and LSST can detect $\sim$ 42 galaxies arcmin$^{-2}$. The effective galaxy number density of CSS-OS is only 1/3 of LSST. But CSS-OS as a space-based project has advantages for the detection of small size galaxies. We can find that there is $\sim$ 1.5 arcmin$^{-2}$ additional galaxies detected only by CSS-OS comparing to LSST for the galaxy samples we selected. The feature of the "CSS-OS only" galaxies is shown in Fig. \ref{fig:cssos_lsst_mag_re} top figure, we can see that almost all these galaxies have a smaller size (r$_e <$ 0.3 ") and not low surface brightness. While there are $\sim$  31 arcmin$^{-2}$ additional galaxies which can be detected only by LSST comparing to CSS-OS. We show the "CSS-OS not" samples in Fig. \ref{fig:cssos_lsst_mag_re} bottom figure and the galaxy samples can't be detected by CSS-OS for low galaxy surface brightness and the little exposure time of CSS-OS. 

In addition, the depth (point source, 5$\sigma$) of CSS-OS deep survey is similar to that of LSST. As a result, we compare CSS-OS deep survey with LSST for discussing the advantage of the space-based telescope. The space-based telescopes have the feature of low background light and high spatial resolution, so a small aperture telescope with little exposure time can get similar depth (for a point source) with that of a large aperture ground-based telescope with more time.
Fig. \ref{fig:cssos_deep_mag_re} shows the detected galaxy feature of CSS-OS deep survey and LSST. From this figure, we find that LSST has an advantage in the faint galaxy detection due to its large aperture and more exposure time. Over 93$\%$ of the "LSST only" detected galaxies are faint surface brightness galaxies with AB magnitude $>$ 24.5 AB magnitude arcsec$^{-2}$. CSS-OS deep survey seems to be good at detecting small size galaxies for its high spatial resolution. Over 90$\%$ of the "CSS-OS deep survey only" detected galaxies have small size $r_e <$ 0.3".

Both CSS-OS galaxy samples and LSST galaxy samples are mainly constituted with late-type galaxies, the ratio of which reaches over 80$\%$ by morphological information. Meanwhile,  small size galaxies are the main components of the detected galaxy samples. Both CSS-OS and LSST are good at detecting small galaxies, and there are about 50$\%$ galaxies which have small effective radius $r_e <$ 0.3" for the galaxies detected by the two projects. While for the CSS-OS deep survey, the small galaxies ratio exceeds 60$\%$.

From CSS-OS galaxy samples, we can conclude that CSS-OS galaxy observation focus on the galaxies z $<$ 0.8. Due to the limitation of accumulated exposure time, the galaxies number density of CSS-OS is much smaller than that of LSST. But for small size galaxy detection, CSS-OS can discover unique samples which have a small size for its high spatial resolution with respect to what can be achieved from the ground (LSST). Moreover, as the total exposure time increases, the number of these unique samples becomes larger. While comparing with other space telescopes (EUCLID and WFIRST), CSS-OS still has an advantage in image quality due to its off-axis optical system design ($R_{80}$ of Euclid and WFIRST both exceed 0.2").

\section{ACKNOWLEDGEMENTS}
\label{sec:acknowledgements}
XZ acknowledges the support of National Natural Science Foundation of China (Grant No. U1731127) and the Opening Project of Key Laboratory of Computational Astrophysics, National Astronomical Observatories, Chinese Academy of Sciences. LC acknowledges the support by the Open Research Fund of Key Laboratory of Space Utilization, Chinese Academy of Sciences(No. LSU-KFJJ-2018-09). We thank Prof. Hu Zhan for the discussion about methods of image simulation and data analysis, Youhua Xu and Qiao Wang for the discussion about the algorithms, and Yuan Liu and Hongyuan Cai for revising the manuscript.


\nocite{*}
\bibliographystyle{spr-mp-nameyear-cnd}
\bibliography{CSSOS}

\label{lastpage}
\end{document}